\documentclass[conference]{IEEEtran}
\IEEEoverridecommandlockouts
\usepackage{cite}
\usepackage{amsmath,amssymb,amsfonts}
\usepackage{algorithmic}
\usepackage{graphicx}
\usepackage{textcomp}
\usepackage{xcolor}
\usepackage{float}
\usepackage{placeins}
\usepackage{booktabs}
\usepackage{subfigure}
\usepackage{amsmath}
\usepackage{epstopdf}
\usepackage{epsfig}
\usepackage[ruled]{algorithm2e}
\usepackage{multirow}
\usepackage{hyperref}
\usepackage{xcolor}
\usepackage{colortbl} 
\usepackage{hhline} 
\usepackage{longtable}
\usepackage{lscape}
\usepackage{float}
\usepackage{mathtools}
\hypersetup{hidelinks}
\def\BibTeX{{\rm B\kern-.05em{\sc i\kern-.025em b}\kern-.08em
		T\kern-.1667em\lower.7ex\hbox{E}\kern-.125emX}}
\begin{document}

\title{Handover-Aware Trajectory Planning for Cellular-Connected UAVs under STL Specifications and URLLC Constraints}

	\author{Yuqi Ping, Tianhao Liang, Bofeng Zheng, Bingyan Xie, and Tingting Zhang
	\thanks{Yuqi Ping, Tianhao Liang, Bofeng Zheng, and Tingting Zhang are with the Guangdong Provincial Key Laboratory of Space-Aerial Networking and Intelligent Sensing, Harbin Institute of Technology, Shenzhen, P. R. China (e-mail: pingyq@stu.hit.edu.cn; liangth@hit.edu.cn; 25s052015@stu.hit.edu.cn; zhangtt@hit.edu.cn).}
	\thanks{Bingyan Xie is with the Department of Electronic Engineering, Shanghai Jiao Tong University, Shanghai 200240, China (e-mail: bingyanxie@sjtu.edu.cn).}}

\maketitle

\begin{abstract}
	This paper investigates handover-aware trajectory planning for cellular-connected UAVs executing mission-centric tasks under ultra-reliable low-latency communication (URLLC) constraints. Signal temporal logic (STL) provides a formal specification layer for translating mission semantics into time-bounded trajectory requirements, while finite-blocklength URLLC feasibility characterizes reliable command-and-control (C2) links with serving base stations (BSs). We formulate a joint planning problem that optimizes the UAV trajectory, STL mission satisfaction, serving-BS association, and handover behavior. To solve this mixed discrete-continuous problem, we adopt and integrate a Logic Network Flow (LNF) based STL reformulation with B\'ezier-parameterized motion, disk-shaped URLLC service regions, and binary association variables, so that the resulting mixed-integer quadratically constrained formulation can be solved by standard branch-and-bound solvers. Numerical simulations over a library of STL missions show that the proposed planner can execute different mission specifications under the same cellular map while maintaining URLLC feasibility. The results further reveal how mission timing, handover-aware association, and finite-blocklength stringency jointly affect trajectory shape, serving margin, and computational complexity.
	
\end{abstract}


\section{Introduction}
\subsection{Background and Motivation}

Recent advances in artificial intelligence (AI) and embodied autonomy are enabling low-altitude UAVs to evolve from remotely operated platforms into mission-oriented autonomous agents \cite{zhou2026lsailargesmallai}. This transition is expanding the operational role of UAVs from simple flight execution to task-level decision making in applications such as data collection \cite{11177503,liang2024scc}, precision agriculture \cite{10531194}, forest firefighting \cite{ping2025multimodal}, localization and sensing services \cite{liang2023age,liang2026uavdetection}, communication support \cite{liang2024uavaided}, and low-altitude logistics \cite{zhou2026delivery}. In these scenarios, a UAV is no longer required only to reach a destination, but also to visit prescribed regions, respect temporal ordering and safety constraints, and execute sensing or service actions at appropriate mission stages. These coupled spatial and temporal requirements call for mission specifications that can be interpreted by human operators and agentic AI systems, converted into executable planning constraints, and monitored during execution. Signal temporal logic (STL) provides a natural formal language for this purpose, as it can encode region-reaching, temporal ordering, persistence, and safety requirements over mission trajectories \cite{maler2004monitoring,belta2017formal,belta2019formal,raman2014model,ping2026nlstl}.

Unlike many ground-robot or closed-loop task-planning settings where communication is not explicitly modeled, low-altitude UAV missions are often executed over larger areas and under stricter operational supervision. As autonomy increases, a UAV is expected to maintain command-and-control (C2) connectivity with ultra-reliable low-latency communication (URLLC) support via serving base stations (BSs) throughout task execution \cite{popovski2018wireless,bennis2018ultrareliable}. Such cellular connectivity can provide wide-area coverage and beyond-visual-line-of-sight support for these operations \cite{lin2018sky,muruganathan2018overview,zeng2019accessing,zeng2019cellularconnected,zhang2018cellular}. However, mobility across heterogeneous cells tightly couples serving-cell association and handover decisions with trajectory planning. Frequent or poorly planned handovers may degrade communication continuity, incur signaling and synchronization overhead, trigger ping-pong effects, increase latency, and compromise the reliability of C2 links \cite{madelkhanova2022optimization,du2025handover}.

The coexistence of STL-specified mission requirements and handover-sensitive C2 connectivity exposes a gap between formal mission planning and cellular-connected UAV trajectory design. Mission-level planners encode task objectives and temporal requirements, whereas handover-aware trajectory planners characterize feasible UAV motion under connectivity constraints. Such temporal requirements directly affect the sequence of serving cells encountered along the trajectory. A decoupled design may therefore produce trajectories that satisfy task logic while ignoring serving-cell feasibility, or trajectories that maintain connectivity while violating mission timing and logical dependencies. This gap motivates a unified handover-aware trajectory planning formulation that jointly optimizes UAV motion, STL mission satisfaction, base-station association, and handover behavior, so that task completion and reliable network support are treated as coupled requirements rather than separate design stages.

\subsection{Related Works}

Mission-centric UAV planning requires formal task specifications that can represent not only geometric goals but also temporal and logical mission semantics. Temporal logic has been applied to reactive mission planning, robot deployment, multi-robot coordination, and motion planning with formal specifications \cite{kressgazit2009temporal,kloetzer2010automatic,plaku2016motion,kantaros2020stylus}. For dynamical systems, STL and related temporal logics describe time-bounded requirements over continuous-valued signals, making them suitable for region visiting, sequencing, persistence, deadline, and safety constraints \cite{maler2004monitoring,fainekos2009robustness,donze2013efficient,belta2017formal,belta2019formal}. Such specifications have been embedded in mixed-integer optimization, model predictive control, and reactive synthesis formulations \cite{raman2014model,raman2015reactive,wolff2014optimization,sun2022multiagent}. Language-grounded mission specification further supports translating human-understandable commands into machine-checkable temporal-logic formulas, making STL a candidate language for UAV mission specifications in agentic planning pipelines \cite{dzifcak2009translating,finucane2010ltlmop,ping2026nlstl}.

From an optimization perspective, STL mission planning often relies on mixed-integer encodings whose size and relaxation quality can become problematic over long horizons \cite{kurtz2022scalable,kurtz2022mixed}. Logic network flow (LNF) encodes temporal-logic satisfaction through network-flow constraints and can provide tighter convex relaxations than conventional tree-based encodings \cite{lin2025lnf}, while convex and mixed-integer motion-planning formulations provide compatible trajectory parameterizations and motion-feasibility tools \cite{richards2002aircraft,mellinger2011minimum,deits2015computing,deits2015efficient,schulman2014motion,marcucci2023motion,marcucci2024shortest}. Nevertheless, these optimization tools are typically used for task-and-motion feasibility, and do not by themselves capture serving-BS selection, handover counting, or finite-blocklength communication constraints during mission execution.

Handover-aware cellular-connected UAV trajectory planning has been studied in parallel as an enabling technology for practical low-altitude autonomy. Early, tutorial, and survey works identify UAV communication opportunities, air-to-ground channel characteristics, mobility-induced coverage variations, and cellular-connected UAV challenges \cite{hayat2016survey,zeng2016wireless,mozaffari2019tutorial,zeng2019accessing,zeng2019cellularconnected,fotouhi2019survey,li2019uavcommunications}. Standardization and measurement studies further show that aerial UEs experience distinct propagation, interference, and mobility behaviors in cellular systems \cite{lin2018sky,muruganathan2018overview,khawaja2019survey,amorim2018measured,geraci2018understanding}. Network architectures and resource orchestration for UAV-assisted and air-ground integrated networks have also been investigated \cite{kawamoto2019toward,cheng2018airground}. Machine-learning-based connectivity and security mechanisms have been studied for cellular-connected UAVs \cite{challita2019machine}. Building on these foundations, trajectory planning for cellular-connected UAVs has been studied under connectivity, outage, energy, sensing, and handover constraints. A representative line of work formulates connectivity-constrained UAV planning as a graph or waypoint-search problem, where graph edges or waypoints are refined under distance, outage, battery, propulsion-energy, or handover-related constraints \cite{zhang2018cellular,zhang2019trajectory,im2025trajectory,yang2021efficient,du2025handover}. Continuous-time GCS-based trajectory optimization has also been used to jointly optimize URLLC-supported UAV motion and BS association under handover costs \cite{ping2025handoverurllcgcs}. Other studies consider UAV-enabled sensing, localization, integrated sensing/communication, resource allocation, and data collection, where trajectory, communication, and sensing schedules must be coordinated \cite{liang2023age,liang2024uavaided,luo2022ris,liang2024scc,zeng2026deep,11177503,liang2026uavdetection}. Satellite--UAV non-terrestrial network architectures further highlight the need for seamless localization and communication support across heterogeneous aerial networks \cite{liang2024seamless}. These studies provide important tools for handover-aware mobility control, but they usually focus on destination-oriented or waypoint-level tasks, often decomposing BS selection and trajectory refinement, and therefore cannot directly represent logical mission requirements such as precedence, return, deadline, and dwell constraints.

URLLC-aware UAV communication and control provide a further set of network constraints that are directly relevant to mission execution. URLLC studies emphasize that reliability, latency, packet size, access delay, and risk-sensitive network design must be treated jointly in mission-critical wireless systems \cite{popovski2018wireless,bennis2018ultrareliable,parvez2018survey,sutton2019enabling,she2021tutorial}. The finite-blocklength regime further couples packet reliability, latency, blocklength, and received signal quality \cite{polyanskiy2010channel,durisi2016toward,shirvanimoghaddam2019short}. For aerial links, multi-connectivity and power-control mechanisms have been studied to support URLLC-enabled UAV communication \cite{salehi2022ultra,yang2021power}. Recent works further consider event-triggered and self-triggered control for remote UAV systems, as well as communication-control co-design for trajectory tracking, distributed formation, and energy-efficient data collection under wireless uncertainty \cite{11310228,11174804,ping2024formation,liang2024scc}. Low-altitude intelligent networking and UAV swarm systems have also motivated stronger coordination among perception, communication, and autonomous decision making \cite{ping2025multimodal,lei2026secure,zhou2026delivery}. However, these studies usually treat reliability and latency as link- or control-layer requirements under a prescribed task structure, rather than as constraints that jointly shape mission timing, serving-cell selection, and handover-aware motion planning.

Despite these advances, three issues remain unresolved for mission-centric cellular-connected UAV autonomy. First, formal mission-planning methods can express temporal and logical task requirements, but they usually abstract away the serving-cell sequence and communication feasibility that determine whether the mission can be executed over a real network. Second, cellular-connected UAV and URLLC studies optimize trajectory, resource allocation, and link reliability, but the task layer is often reduced to destination reaching, waypoint visiting, or preassigned operation phases. Third, handover-aware designs rarely coordinate mission phases, serving-cell selection, and handover behavior within a single optimization model. These limitations motivate a unified handover-aware planning framework in which STL mission satisfaction, finite-blocklength URLLC feasibility, BS association, handover behavior, and UAV motion are planned jointly under coupled mission and network constraints.

\subsection{Main Contributions}

This paper aims to enable mission-centric trajectory planning for cellular-connected UAVs by jointly considering STL mission semantics, finite-blocklength URLLC feasibility, serving-BS association, and handover behavior. The main contributions are summarized as follows.
\begin{itemize}
	\item We first establish a handover-aware cellular-connected UAV planning framework for STL-specified missions. In this framework, the mission layer encodes temporal task requirements such as ordered inspection, disjunctive service selection, deadline-constrained inspection, data-upload, and return, while the network layer models URLLC-supported serving regions, BS association, and handover events.
	\item Next, we formulate a joint mission-network-motion optimization problem over a fixed mission horizon. The decision variables include B\'ezier control points, STL predicate satisfaction variables, serving-BS association binaries, and handover indicators, and the objective balances motion regularity and handover cost subject to STL satisfaction, UAV motion constraints, and finite-blocklength URLLC feasibility.
	\item We then adopt and integrate an LNF-based STL reformulation with B\'ezier-parameterized motion and disk-shaped URLLC service regions. This construction maps the coupled STL, association, handover, and motion planning problem into a mixed-integer quadratically constrained program.
	\item Finally, we perform numerical simulations on a four-mission library and network-sensitivity studies. The results demonstrate that the same planner can execute different STL specifications under one cellular map, and they quantify how handover-aware association and finite-blocklength stringency affect serving margin, feasible mission execution, and computational footprint.
\end{itemize}

\begin{figure}[h]
	\centering
	\includegraphics[width=\columnwidth]{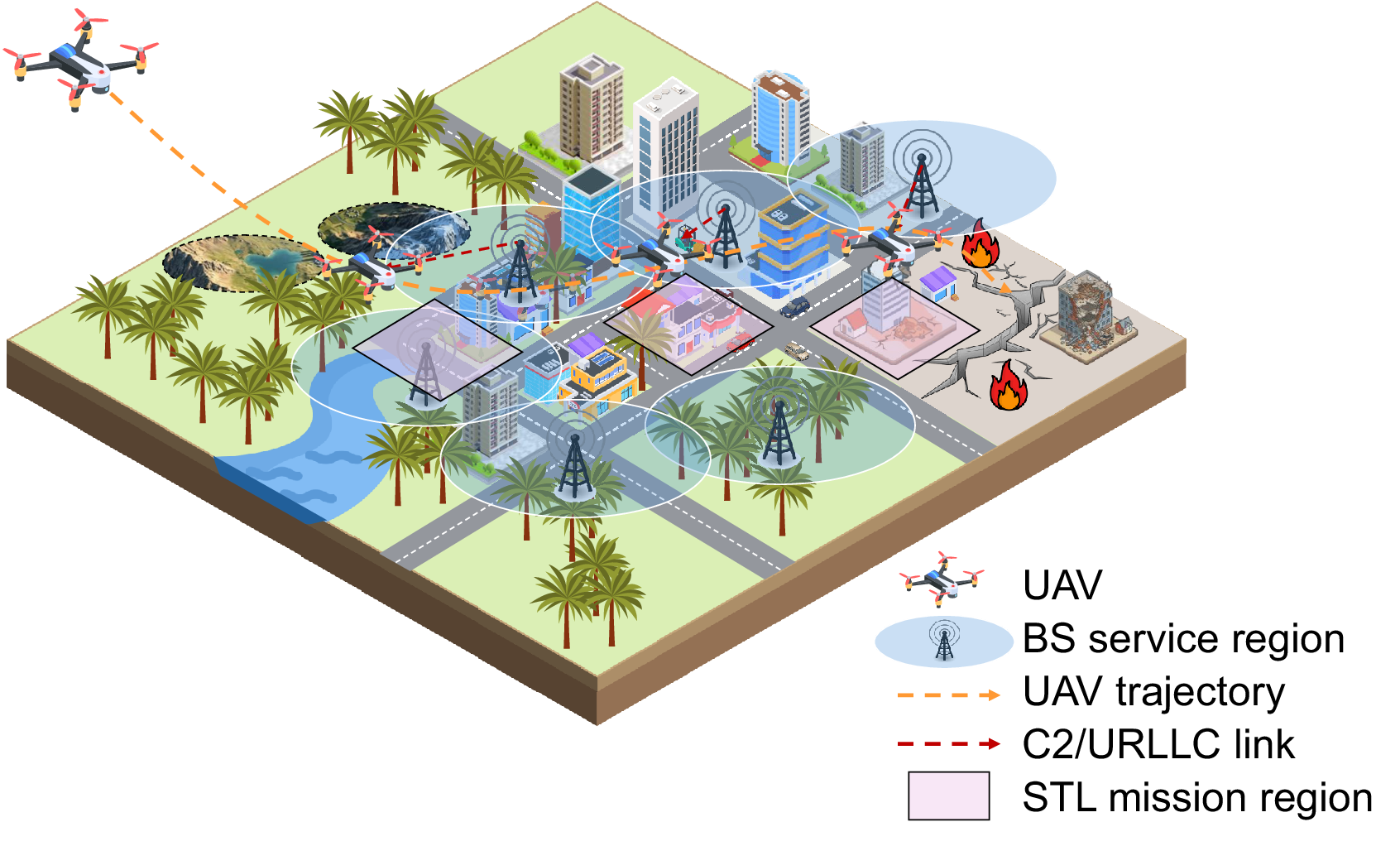}
	\vspace{-0.6em}
	\caption{System model of the handover-aware cellular-connected UAV planning framework.}
	\label{fig:system model}
\end{figure}

\section{System Model}
\label{sec:system_model}
\subsection{System Overview}
As illustrated in Fig.~\ref{fig:system model}, we consider a cellular-connected UAV that performs a mission under cellular coverage and URLLC constraints. The feasible airspace is denoted by $\mathcal{Q}\subset\mathbb{R}^3$, and the UAV flies at a fixed altitude $H$. Different from conventional point-to-point trajectory planning, the task is described by a mission specification that states what the UAV should accomplish and when it should be accomplished. Let $\mathcal{R}=\{\mathcal{R}_1,\ldots,\mathcal{R}_{N_R}\}$ denote a set of ground-plane semantic mission regions with $\mathcal{R}_j\subset\mathbb{R}^2$, where $N_R$ is the number of mission regions. These regions may include, for example, inspection areas, monitoring zones, relay-support locations, recovery regions, and emergency-service regions. These semantic regions and their temporal relations may be obtained from a task database, a semantic map, or a higher-level language-to-logic module. This paper takes the resulting machine-interpretable predicates and temporal constraints as inputs and focuses on executing them through handover-aware UAV planning.

Let $\mathcal{B}=\{1,\ldots,M\}$ denote the set of terrestrial BSs, where $M$ is the number of BSs. During the mission, the UAV is associated with one serving BS at each time instant and may hand over between BSs along its trajectory. The C2 link between the UAV and the serving BS must satisfy URLLC requirements throughout the mission horizon. Therefore, the planner has to determine the UAV trajectory, serving-BS association, handover schedule, and mission satisfaction jointly. This setting requires coordination between the task layer and the communication layer. Formal mission specifications define the required mission behavior, while the cellular infrastructure provides the reliable C2 support needed during execution.

\subsection{UAV Motion Model}
We model the UAV motion by a continuous-time, fixed-altitude trajectory over the finite mission horizon $[0,T]$. Let $\boldsymbol{q}(t)=[x(t),\,y(t),\,z(t)]^\top\in\mathbb{R}^3$ denote the UAV position, let $\boldsymbol{q}_s$ denote the initial position, and let $\boldsymbol{v}(t)=\dot{\boldsymbol{q}}(t)$ denote its velocity. The flight altitude is fixed to $z(t)=H$, and the planar trajectory is required to be continuous and piecewise smooth over the mission horizon. The UAV is subject to the maximum-speed constraint
\begin{equation}
	\|\boldsymbol{v}(t)\|_2 \le v_{\max}, \qquad \forall\,t\in[0,T],
\end{equation}
where $v_{\max}$ is the maximum flight speed. The UAV position satisfies $[x(t),y(t),H]^\top\in\mathcal{Q}$ throughout the mission horizon.

\vspace{-0.3em}
\subsection{UAV-to-BS Communication Model}

Following the UAV-to-ground communication model in \cite{al2014optimal}, we characterize each UAV--BS link by a combination of line-of-sight (LoS) and non-line-of-sight (NLoS) components. For each BS $i \in \mathcal{B}$, its three-dimensional position is
\begin{equation}
	\boldsymbol{s}_{i} = [x_{i}^{\mathrm{BS}},\,y_{i}^{\mathrm{BS}},\,z_{i}^{\mathrm{BS}}]^\top \in \mathbb{R}^3,
\end{equation}
where $x_i^{\mathrm{BS}}$ and $y_i^{\mathrm{BS}}$ represent the ground-plane coordinates and $z_i^{\mathrm{BS}}$ denotes the antenna height of BS $i$.

At time $t$, the 3D link distance and elevation angle between the UAV and BS $i$ are given by
\begin{align}
	d_{i}(t) &= \big\|\,\boldsymbol{q}(t)-\boldsymbol{s}_{i}\,\big\|,\\
	\theta_{i}(t) &= \arctan\!\Big(\frac{z(t)-z_{i}^{\mathrm{BS}}}{\sqrt{(x(t)-x_{i}^{\mathrm{BS}})^2+(y(t)-y_{i}^{\mathrm{BS}})^2}}\Big).
\end{align}

Under this UAV-to-ground channel model, the elevation-dependent LoS and NLoS probabilities are modeled as
\begin{align}
	p^{\mathrm{LoS}}_i(t) &= \frac{1}{1 + \alpha_{\mathrm{env}}\,
	\exp\!\big(-\beta_{\mathrm{env}}(\theta_i(t)-\alpha_{\mathrm{env}})\big)},\\
	p^{\mathrm{NLoS}}_i(t) &= 1 - p^{\mathrm{LoS}}_i(t),
\end{align}
where $\alpha_{\mathrm{env}}$ and $\beta_{\mathrm{env}}$ are environment-dependent parameters determined by the type of terrain.

Accordingly, the average path loss between the UAV and BS $i$ can be expressed as
\begin{equation}
	L_{i}(t) = p^{\mathrm{LoS}}_i(t)\,L_{i}^{\mathrm{LoS}}(t)
	+ p^{\mathrm{NLoS}}_i(t)\,L_{i}^{\mathrm{NLoS}}(t),
\end{equation}
where the LoS and NLoS components are respectively given by
\begin{align}
	L_{i}^{\mathrm{LoS}}(t)   &= \Big(\tfrac{4\pi f_c\, d_{i}(t)}{c}\Big)^2 \zeta^{\mathrm{LoS}},\\
	L_{i}^{\mathrm{NLoS}}(t) &= \Big(\tfrac{4\pi f_c\, d_{i}(t)}{c}\Big)^2 \zeta^{\mathrm{NLoS}},
\end{align}
with $f_c$ denoting the carrier frequency, $c$ the speed of light, and $\zeta^{\mathrm{LoS}}$, $\zeta^{\mathrm{NLoS}}$ the excessive path loss factors corresponding to the LoS and NLoS cases, respectively.

Given the BS transmit power $P_i$, receiver antenna gain $g_r$, and receiver noise power $\sigma^2$, the received signal-to-noise ratio (SNR) associated with BS $i$ at the UAV position $\boldsymbol{q}(t)$ is
\begin{equation}
	\gamma_i(\boldsymbol{q}(t)) = \frac{g_r\,P_{i}}{L_{i}(t)\,\sigma^2}.
\end{equation}

\subsection{Finite-Blocklength URLLC Feasibility}

We impose a URLLC requirement on the UAV--BS C2 link in terms of latency, reliability, and coding rate. 
Let $L$ denote the end-to-end latency budget and $\varepsilon_{\max}$ the target decoding error probability. 
Given system bandwidth $B$ and transmission duration $\tau$, the channel blocklength is
\begin{equation}
	N_{\mathrm{fb}} = B\tau, \qquad 0< \tau \le L, \qquad 0< \varepsilon \le \varepsilon_{\max},
	\label{eq:blocklength}
\end{equation}
where $\varepsilon$ is the packet error probability, while $\tau\le L$ and $\varepsilon\le\varepsilon_{\max}$ impose the latency and reliability requirements, respectively.

Under the normal approximation for finite blocklength coding \cite{durisi2016toward}, the achievable per-channel-use rate is
\begin{equation}
	R(\gamma,N_{\mathrm{fb}},\varepsilon)
	\approx \log_2(1+\gamma)\,-\,\sqrt{\frac{V(\gamma)}{N_{\mathrm{fb}}}}\,Q^{-1}(\varepsilon)\,+\,\frac{\log_2 N_{\mathrm{fb}}}{2N_{\mathrm{fb}}},
	\label{eq:Rfb}
\end{equation}
where the channel dispersion is
\begin{equation}
	V(\gamma) = \frac{\gamma(\gamma+2)}{(1+\gamma)^2}\,(\log_2 e)^2,
\end{equation}
$\gamma$ is the receive SNR, and $Q^{-1}(\cdot)$ is the inverse Gaussian $Q$-function.

Let $R_{\mathrm{req}}$ denote the required coding rate for the C2 link. The URLLC feasibility condition is
\begin{equation}
	R(\gamma,N_{\mathrm{fb}},\varepsilon_{\max}) \,\ge\, R_{\mathrm{req}}.
	\label{eq:urlcc-feasible}
\end{equation}
For a given $(N_{\mathrm{fb}},\varepsilon_{\max})$, define the minimum required SNR as
\begin{equation}
	\gamma_{\min}(N_{\mathrm{fb}}) \,\triangleq\, \inf\left\{\gamma \ge 0:\,
	R(\gamma,N_{\mathrm{fb}},\varepsilon_{\max}) \ge R_{\mathrm{req}}\right\}.
\end{equation}

For the selected blocklength, let $\gamma_{\min}\triangleq\gamma_{\min}(N_{\mathrm{fb}})$. Each BS induces a URLLC-feasible region
\begin{equation}
	\widetilde{\mathcal{C}}_i \,\triangleq\, \big\{\, \boldsymbol{q} \in \mathcal{Q}:\, \gamma_i(\boldsymbol{q}) \ge \gamma_{\min} \,\big\},
\end{equation}
where $\gamma_i(\boldsymbol{q})$ denotes the SNR from BS $i$ evaluated at an arbitrary UAV position $\boldsymbol{q}$. Thus, the optimized serving association $b(t)$ must satisfy $\gamma_{b(t)}(\boldsymbol{q}(t))\ge\gamma_{\min}$ over the mission horizon.

\subsection{STL-Based Mission Specification}
\label{subsec:mission_stl}

Consistent with the system overview, the semantic mission regions $\mathcal{R}$ and their temporal relations are treated as machine-interpretable inputs. This subsection formalizes these inputs as a predicate library and a bounded-time STL formula. Predicate and formula generation from task databases, semantic maps, or natural-language commands can be handled by upstream mission-understanding modules \cite{ping2026nlstl}, while this paper focuses on planning with the resulting predicate library and temporal formula.

Let $\Pi=\{\pi_1,\ldots,\pi_{N_\pi}\}$ denote the set of atomic task predicates defined over the UAV trajectory $\boldsymbol{q}$ on $[0,T]$. We assume that each semantic mission region can be represented by a polygonal ground-plane footprint given by
\begin{equation}
	\mathcal{R}_j=\{\boldsymbol p\in\mathbb{R}^2:
	\boldsymbol{a}_{j,r}^{\top}\boldsymbol p\le \kappa_{j,r},\ r=1,\ldots,F_j\},
	\label{eq:mission_region_poly}
\end{equation}
where $\boldsymbol p=[x,y]^\top$, $F_j$ is the number of polygon facets, and $(\boldsymbol{a}_{j,r},\kappa_{j,r})$ defines the $r$th half-space boundary of $\mathcal{R}_j$. The corresponding predicate $\pi_j$ is satisfied when the horizontal UAV position $\boldsymbol{q}_{xy}(t)=[x(t),y(t)]^\top$ lies in $\mathcal{R}_j$. These predicates are used to encode mission semantics such as reaching an inspection area, staying in a monitoring zone, entering a recovery region, or visiting a service region within a prescribed time window, and they serve as atomic mission conditions in the STL formula.

The STL syntax specifies how atomic predicates are recursively composed into Boolean and temporal requirements over the mission horizon. Following the standard STL syntax \cite{maler2004monitoring}, a bounded-time formula is generated by
\begin{equation}
\begin{aligned}
	\varphi ::= {}& \pi \,|\, \neg\varphi
	\,|\, \varphi_1\wedge\varphi_2
	\,|\, \varphi_1\vee\varphi_2 \\
	&\,|\, \Box_{[t_1,t_2]}\varphi
	\,|\, \Diamond_{[t_1,t_2]}\varphi
	\,|\, \varphi_1\,\mathcal{U}_{[t_1,t_2]}\,\varphi_2,
\end{aligned}
	\label{eq:stl_syntax_general}
\end{equation}
where $\pi\in\Pi$, $0\le t_1\le t_2\le T$, and $\Box$, $\Diamond$, and $\mathcal{U}$ denote always, eventually, and until operators, respectively. We write $\boldsymbol{q}(\cdot)\models\varphi$ when the UAV trajectory satisfies the formula.

Let $\varphi_{\mathrm{mis}}$ denote the bounded-time STL mission formula constructed from $\Pi$ over horizon $T$. Different mission cases are obtained by changing the predicate library and the temporal composition of $\varphi_{\mathrm{mis}}$, while the handover-aware planning architecture remains unchanged.

\section{Problem Formulation}
\label{sec:problem_formulation}

Given a predicate library $\Pi$ and a bounded-time STL mission formula $\varphi_{\mathrm{mis}}$ over a fixed mission horizon $T$, the planner jointly determines the UAV trajectory and serving-BS association. The trajectory must satisfy $\varphi_{\mathrm{mis}}$ through the mission predicates defined over the UAV motion. Let $b(t)\in\mathcal{B}$ denote the serving BS index at time $t$, where $b(\cdot)$ is piecewise-constant and right-continuous.

The handover time set is defined as
\begin{equation}
	\mathcal{T}_{\mathrm{ho}}\triangleq\{\,t\in(0,T]:\, b(t^-)\neq b(t)\,\},
\end{equation}
and $N_{\mathrm{ho}}\triangleq|\mathcal{T}_{\mathrm{ho}}|$ denotes the total number of handovers. Thus, handover awareness enters through the time variation of $b(t)$. Maintaining the same feasible serving BS over consecutive time intervals can reduce service transitions, but it may require a larger motion effort or a smaller serving margin than always selecting the BS with the largest instantaneous SNR. We therefore introduce a nonnegative motion-regularization functional $\mathcal{J}_{\mathrm{m}}(\boldsymbol{q})$ to penalize inefficient or abrupt UAV motion, and combine it with the handover count in the objective. With nonnegative weight $w_{\mathrm{ho}}$ trading off service transitions against motion regularization, the resulting handover-aware STL mission-planning problem under URLLC constraints is formulated as
\begin{subequations}\label{prob:P1}
	\begin{align}
		\text{(P1)}\quad 
		&\min_{\substack{\boldsymbol{q},\,\boldsymbol{v},\\ b}}
		\ \ w_{\mathrm{ho}}\,N_{\mathrm{ho}}\ +\ \mathcal{J}_{\mathrm{m}}(\boldsymbol{q}) \notag\\[2pt]
		\text{s.t.}\quad 
		& \dot{\boldsymbol{q}}(t)=\boldsymbol{v}(t), \hfill \forall t\in[0,T], \label{eq:dyn}\\
		& \|\boldsymbol{v}(t)\|_2\le v_{\max}, \hfill \forall t\in[0,T], \label{eq:speed}\\
		& \boldsymbol{q}(0)=\boldsymbol{q}_s, \label{eq:initial}\\
		& z(t)=H, \hfill \forall t\in[0,T], \label{eq:altitude}\\
		& \boldsymbol{q}(t)\in\mathcal{Q},\quad b(t)\in\mathcal{B}, \hfill \forall t\in[0,T], \label{eq:feasible}\\
		& \gamma_{\,b(t)}\big(\boldsymbol{q}(t)\big)\ \ge\ \gamma_{\min}, \hfill \forall t\in[0,T], \label{eq:urlcc}\\
		& \boldsymbol{q}(\cdot) \models \varphi_{\mathrm{mis}}. \label{eq:mission_sat}
	\end{align}
\end{subequations}

Problem P1 contains three coupled layers. Constraints \eqref{eq:dyn}--\eqref{eq:feasible} describe UAV motion, fixed-altitude flight, feasible airspace, and BS-index feasibility. Constraint \eqref{eq:urlcc} enforces URLLC-supported C2 connectivity through the selected serving BS at every time instant. Constraint \eqref{eq:mission_sat} enforces the STL mission layer by requiring the physical trajectory to satisfy the temporal mission formula. Therefore, the decision variables must determine not only a bounded-speed and URLLC-connected flight path, but also when each mission predicate becomes true, whether the temporal relations are satisfied, and how serving-BS handovers are scheduled during mission execution.

It is worth noting that the objective of P1 contains the handover count $N_{\mathrm{ho}}$, which is induced by changes in the discrete-valued serving-BS function $b(t)$. The constraints also involve continuous UAV motion, discrete serving-BS association, and STL satisfaction requirements that introduce logical dependencies over time. Therefore, P1 is a mixed discrete-continuous, nonconvex, and logic-constrained optimal control problem, for which a direct application of standard convex optimization methods is not available. This motivates a mixed-integer optimization reformulation.
\section{LNF-Based Reformulation with B\'ezier-Parameterized Motion}
\label{sec:lnf_bezier}
This section converts P1 into a mixed-integer quadratically constrained program by transcribing the UAV motion with B\'ezier control points, imposing URLLC-constrained BS association and handover variables, and encoding the STL mission layer through LNF constraints \cite{lin2025lnf}. In this formulation, B\'ezier control points jointly determine motion feasibility, network feasibility, and mission-predicate truth values. The resulting problem can be handled by standard branch-and-bound algorithms that solve convex quadratic or conic relaxations at the search nodes.

\subsection{B\'ezier Motion Parameterization}
\label{subsec:bezier_motion}

The continuous trajectory constraint in P1 is difficult to impose directly together with discrete association and logic variables. We therefore use fixed-duration B\'ezier segments, which preserve a continuous trajectory representation while providing control points that can be constrained through convex geometry.

We represent the UAV trajectory by $N$ cubic B\'ezier segments, each with fixed duration $\Delta t$, which yields the mission horizon $T=N\Delta t$. Since the UAV flies at the fixed altitude $H$, we parameterize only its horizontal trajectory, denoted by $\boldsymbol r(t)\in\mathbb{R}^2$, and recover the three-dimensional position as $\boldsymbol q(t)=[\boldsymbol r(t)^\top,H]^\top$. For segment $n$, the planar position curve is
\begin{equation}
	\boldsymbol r_n(\xi)=\sum_{\ell=0}^{3}\boldsymbol r_{n,\ell}B_{\ell}^{(3)}(\xi),\qquad \xi\in[0,1],
	\label{eq:bezier_segment}
\end{equation}
where $\boldsymbol r_{n,\ell}\in\mathbb{R}^2$ are spatial control points and $B_{\ell}^{(3)}(\xi)=\binom{3}{\ell}(1-\xi)^{3-\ell}\xi^{\ell}$ is the cubic Bernstein basis polynomial.
Coverage and predicate constraints are imposed on the four control points of each fixed-duration segment. If all four control points of segment $k$ lie in a convex mission region, the entire B\'ezier segment lies in that region by the convex-hull property of B\'ezier curves \cite{marcucci2023motion}.

Adjacent B\'ezier segments must share endpoints so that the concatenated curve remains continuous. We impose $C^0$ continuity between neighboring segments through
\begin{equation}\label{eq:bezier_continuity}
	\boldsymbol r_{n,3}=\boldsymbol r_{n+1,0},
	\qquad n=0,\ldots,N-2.
\end{equation}
The initial point is inherited from \eqref{eq:initial}, where $\boldsymbol q_{s,xy}$ denotes the ground-plane projection of $\boldsymbol q_s$ and $\boldsymbol r_{0,0}=\boldsymbol q_{s,xy}$. For $\ell=0,1,2$, the derivative control points satisfy
\begin{equation}
	\boldsymbol r_{n,\ell}^{(1)}=3(\boldsymbol r_{n,\ell+1}-\boldsymbol r_{n,\ell}).
\end{equation}
The maximum-speed constraint is imposed as the second-order cone constraint
\begin{equation}
	\|\boldsymbol r_{n,\ell}^{(1)}\|_2\le v_{\max}\Delta t,
	\qquad n=0,\ldots,N-1,\ \ell=0,1,2.
	\label{eq:soc_speed_bezier}
\end{equation}
This constraint bounds the segment velocity over the full continuous B\'ezier curve because the derivative curve also satisfies the convex-hull property.

\subsection{URLLC-Constrained BS Association and Handover Encoding}
\label{subsec:association_predicates}

After the motion variables are represented by B\'ezier control points, the finite-blocklength URLLC requirement can be imposed directly on each continuous segment. On the fixed-altitude plane $z=H$, the SNR condition $\gamma_i(\boldsymbol q)\ge\gamma_{\min}$ for BS $i$ is equivalent to the path-loss bound
\begin{equation}
	L_i(\|\boldsymbol p-\boldsymbol c_i\|_2;H)
	\le \bar L_i
	\triangleq \frac{g_r P_i}{\sigma^2\gamma_{\min}},
	\label{eq:path_loss_threshold_disk}
\end{equation}
where $\boldsymbol p=[x,y]^\top$, $\boldsymbol c_i=[x_i^{\mathrm{BS}},y_i^{\mathrm{BS}}]^\top$, and $L_i(r;H)$ denotes the UAV-to-BS path-loss model evaluated at altitude $H$ and horizontal separation $r$ from BS $i$. Since $L_i(r;H)$ increases with $r$ under the fixed-altitude channel model, the largest URLLC-feasible horizontal service radius of BS $i$ is
\begin{equation}
	\rho_i\triangleq
	\sup\{\,r\ge0:\ L_i(r;H)\le \bar L_i\,\}.
	\label{eq:urlcc_radius_def}
\end{equation}
The resulting ground-plane service region is the disk
\begin{equation}
	\mathcal{D}_i=\{\boldsymbol p\in\mathbb{R}^2:
	\|\boldsymbol p-\boldsymbol c_i\|_2\le \rho_i\},
	\label{eq:coverage_disk}
\end{equation}
and any point in $\mathcal{D}_i$ satisfies the finite-blocklength URLLC threshold of BS $i$. This disk representation keeps the association constraints compatible with mixed-integer conic optimization.

Let $\delta_{n,i}\in\{0,1\}$ indicate whether segment $n$ is served by BS $i$. A unique serving BS is selected for every segment.
\begin{equation}
	\sum_{i\in\mathcal{B}}\delta_{n,i}=1,\qquad n=0,\ldots,N-1.
	\label{eq:one_bs_segment}
\end{equation}
The segment is constrained to lie in the selected URLLC disk by enforcing disk membership on its B\'ezier control points.
\begin{equation}
	\|\boldsymbol r_{n,\ell}-\boldsymbol c_i\|_2^2
	\le \rho_i^2+M_i(1-\delta_{n,i}),
	\quad \forall i,n,\ell,
	\label{eq:coverage_big_m}
\end{equation}
where $M_i$ is a sufficiently large inactive constant for BS $i$. Since disks are convex, the convex-hull property of B\'ezier curves implies that \eqref{eq:coverage_big_m} keeps the entire segment inside the selected URLLC disk when $\delta_{n,i}=1$.

Under the fixed-duration segment transcription, the continuous-time serving function $b(t)$ is approximated as piecewise constant over each B\'ezier segment. A binary variable $\eta_n\in\{0,1\}$ records whether a handover occurs between consecutive segments. For all $i\in\mathcal{B}$ and $n=0,\ldots,N-2$, impose
\begin{equation}\label{eq:handover_binary}
\begin{aligned}
	\eta_n&\ge \delta_{n,i}-\delta_{n+1,i},\\
	\eta_n&\ge \delta_{n+1,i}-\delta_{n,i},\\
	\eta_n&\le 2-\delta_{n,i}-\delta_{n+1,i}.
\end{aligned}
\end{equation}
Together with the one-BS constraint in \eqref{eq:one_bs_segment}, these inequalities make $\eta_n$ an exact handover indicator. Specifically, $\eta_n=1$ if segments $n$ and $n+1$ select different serving BSs, and $\eta_n=0$ otherwise. The continuous-time handover count $N_{\mathrm{ho}}$ in P1 is thus transcribed as $\sum_{n=0}^{N-2}\eta_n$.

\subsection{Segment-Level STL Predicate Mapping and LNF Encoding}
\label{subsec:lnf_encoding}

After imposing the motion and network constraints, the B\'ezier trajectory must also satisfy the temporal mission specification. To evaluate the STL formula on this trajectory, the continuous curve is first mapped to segment-level region predicates, and LNF is then used to encode the temporal structure of the mission formula.

Under the fixed-duration segment transcription, each B\'ezier segment corresponds to one logic time step of duration $\Delta t$. The STL formula is evaluated on the segment-level predicate trace induced by the control-point hull of each B\'ezier segment. Following the LNF notation, we collect the truth values of all predicate-time instances induced by $\varphi_{\mathrm{mis}}$ into the vector $\boldsymbol{\chi}^{\varphi}=[\chi_1^{\varphi},\ldots,\chi_{N_{\varphi}}^{\varphi}]^\top\in\{0,1\}^{N_{\varphi}}$, which plays the role of the predicate truth vector $\boldsymbol z^\pi$ in \cite{lin2025lnf}. The constraint family $\mathcal{Z}_{\varphi}(\boldsymbol r)$ maps these Boolean variables to the B\'ezier trajectory.

For a region predicate associated with $\mathcal{R}_j$ and evaluated on segment $k$, let $\chi_{k,j}$ denote the corresponding component of $\boldsymbol{\chi}^{\varphi}$. In this paper, segment $k$ satisfies $\pi_j$ if all four control points of that segment lie in $\mathcal{R}_j$. The implication that $\chi_{k,j}=1$ makes the whole B\'ezier segment lie in $\mathcal{R}_j$ is imposed through the convex-hull property by
\begin{equation}
	\boldsymbol a_{j,r}^{\top}\boldsymbol r_{k,\ell}
	\le \kappa_{j,r}
	+M_{j,r}^{\mathrm{in}}(1-\chi_{k,j}).
	\label{eq:inside_region_mip}
\end{equation}
This constraint is imposed for $r=1,\ldots,F_j$ and $\ell=0,\ldots,3$.
When a negated region predicate is used, the complementary implication must also be represented. Introducing binary outside-facet variables $\xi_{k,j,\ell,r}\in\{0,1\}$, we use
\begin{equation}\label{eq:outside_region_mip}
\begin{aligned}
	&\boldsymbol a_{j,r}^{\top}\boldsymbol r_{k,\ell}
	\ge \kappa_{j,r}+\epsilon_j
	-M_{j,r}^{\mathrm{out}}(1-\xi_{k,j,\ell,r}),\\
	&\sum_{\ell=0}^{3}\sum_{r=1}^{F_j}\xi_{k,j,\ell,r}
	\ge 1-\chi_{k,j},
\end{aligned}
\end{equation}
where the first inequality holds for $r=1,\ldots,F_j$ and $\ell=0,\ldots,3$.
Here $\epsilon_j\ge0$ specifies the boundary convention and $M_{j,r}^{\mathrm{in}}$, $M_{j,r}^{\mathrm{out}}$ are sufficiently large constants over the planar workspace. Hence, $\chi_{k,j}=0$ enforces that at least one control point violates one polygon facet.
Together, \eqref{eq:inside_region_mip} and \eqref{eq:outside_region_mip} define the predicate-consistency family $\mathcal{Z}_{\varphi}(\boldsymbol r)$ used below by linking segment-level truth variables to B\'ezier control points.

The temporal-logic part of the mission is represented by a tuple
\begin{equation}
	\mathcal{F}_{\varphi}=(\mathcal{G}_{\varphi},\mathcal{P},\Pi),
	\qquad
	\mathcal{G}_{\varphi}=(\mathcal{V}_{\mathrm{L}},\mathcal{E}_{\mathrm{L}}),
	\label{eq:lnf_tuple}
\end{equation}
where $\mathcal{G}_{\varphi}$ is a directed acyclic graph with source vertex $v_s$ and target vertex $v_t$, and $\Pi$ is the atomic predicate set introduced in Section~\ref{subsec:mission_stl}. The collection $\mathcal{P}=\{\mathcal{P}_e\}_{e\in\mathcal{E}_{\mathrm{L}}}$ assigns to each logic edge a set of predicate-time literals, possibly negated, that must be satisfied in order for the edge to be traversed. Conjunctive subformulas accumulate such literals on the same logic edge, whereas disjunctive subformulas generate alternative branches. Hence, satisfaction of $\varphi_{\mathrm{mis}}$ is encoded as routing one unit of logical flow from $v_s$ to $v_t$ while respecting the predicate labels carried by the selected edges.

\begin{figure}[t]
	\centering
	\includegraphics[width=0.85\columnwidth]{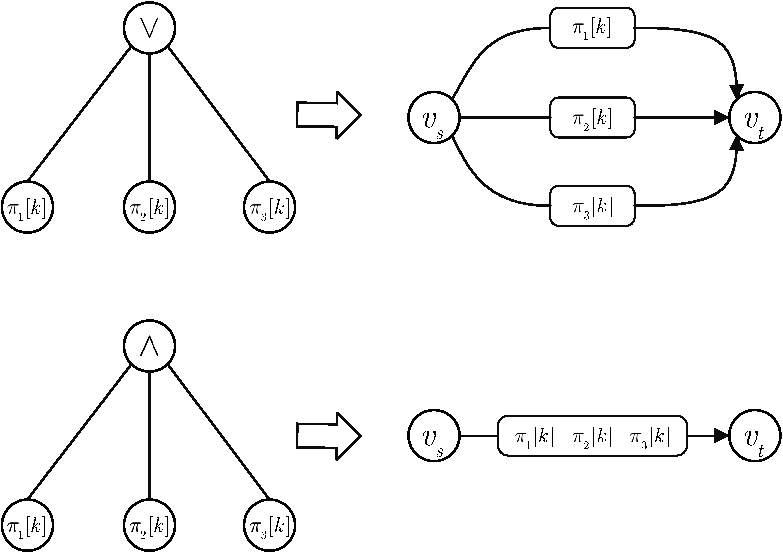}
	\caption{Illustration of translating disjunctive and conjunctive logic nodes into predicate-labeled LNF edges for a generic segment-level index $k$ \cite{lin2025lnf}.}
	\label{fig:lnf_sketches}
\end{figure}

Fig.~\ref{fig:lnf_sketches} illustrates the basic LNF construction for disjunctive and conjunctive logic nodes. A disjunctive node is represented by alternative predicate-labeled edges, whereas a conjunctive node accumulates multiple predicates on the same logic edge. This edge-based view is the reason that STL satisfaction can be represented by a logical flow from the source vertex to the target vertex.

For each $e\in\mathcal{E}_{\mathrm{L}}$, introduce a binary edge-traversal variable $y_e\in\{0,1\}$ and a continuous predicate-flow vector $\boldsymbol\omega_e\in[0,1]^{N_{\varphi}}$. The variable $y_e$ indicates whether logic edge $e$ is used by the logical path, while $\boldsymbol\omega_e$ carries the predicate truth values along that edge. Let $\boldsymbol{\nu}_e^{+},\boldsymbol{\nu}_e^{-}\in\{0,1\}^{N_{\varphi}}$ denote the incidence vectors of the non-negated and negated predicates in $\mathcal{P}_e$, respectively. The edge-label constraints are imposed componentwise as
\begin{equation}\label{eq:lnf_edge_set}
\begin{aligned}
	\boldsymbol\omega_e &\ge y_e\boldsymbol{\nu}_e^{+},
	&&\forall e\in\mathcal{E}_{\mathrm{L}}, \\
	\boldsymbol\omega_e &\le \boldsymbol{1}-y_e\boldsymbol{\nu}_e^{-},
	&&\forall e\in\mathcal{E}_{\mathrm{L}},
\end{aligned}
\end{equation}
where $\boldsymbol{1}$ is the all-one vector. If $y_e=1$, these constraints force every non-negated predicate carried by $e$ to have flow value one and every negated predicate carried by $e$ to have flow value zero. Predicates not appearing in $\mathcal{P}_e$ are unconstrained by that edge. These constraints impose selected-edge-implies-predicate relations. They do not force an edge to become active whenever its label happens to be true, which avoids over-constraining formulas with multiple feasible witness times. The LNF imposes scalar edge traversal and vector predicate-flow conservation.
\begin{equation}\label{eq:lnf_flow}
\begin{aligned}
	\sum_{e\in\mathcal{E}_{v}^{\mathrm{in}}}y_e
	&=\sum_{e\in\mathcal{E}_{v}^{\mathrm{out}}}y_e,
	&&\forall v\in\mathcal{V}_{\mathrm{L}}\setminus\{v_s,v_t\},\\
	\sum_{e\in\mathcal{E}_{v_s}^{\mathrm{out}}}y_e&=1,\\
	\sum_{e\in\mathcal{E}_{v}^{\mathrm{in}}}\boldsymbol\omega_e
	&=\sum_{e\in\mathcal{E}_{v}^{\mathrm{out}}}\boldsymbol\omega_e,
	&&\forall v\in\mathcal{V}_{\mathrm{L}}\setminus\{v_s,v_t\},\\
	\boldsymbol{\chi}^{\varphi}&=\sum_{e\in\mathcal{E}_{v_s}^{\mathrm{out}}}\boldsymbol\omega_e .
\end{aligned}
\end{equation}

The vector flow $\boldsymbol\omega_e$ is an auxiliary copy of the predicate trace propagated along the selected logical path. The last equality in \eqref{eq:lnf_flow} injects the trajectory-induced predicate-time trace into the LNF source. The LNF-enhanced mission-planning problem is obtained by replacing the abstract STL satisfaction constraint $\boldsymbol{q}(\cdot)\models\varphi_{\mathrm{mis}}$ in P1 with the predicate-consistency constraints $\boldsymbol{\chi}^{\varphi}\in\mathcal{Z}_{\varphi}(\boldsymbol r)$ and the network-flow constraints \eqref{eq:lnf_edge_set}--\eqref{eq:lnf_flow}. In the implementation, the LNF flow variables are included in P2 and optimized directly together with the trajectory and BS-association variables.

\subsection{Mixed-Integer Quadratically Constrained Reformulation}
\label{subsec:micp_reformulation}

Once motion, communication, and mission logic have been transcribed, the remaining step is to collect the constraints into one mixed-integer formulation with the objective used in P1. In the current implementation, smoothness is encouraged through the objective rather than imposed as a hard acceleration bound. The motion regularization is made explicit by introducing the second-order B\'ezier difference.

Define the second-order B\'ezier difference
\begin{equation}
	\boldsymbol d_{n,\ell}^{(2)}
	=6(\boldsymbol r_{n,\ell+2}-2\boldsymbol r_{n,\ell+1}+\boldsymbol r_{n,\ell}),
	\quad \ell=0,1.
	\label{eq:bezier_second_difference}
\end{equation}
In the B\'ezier transcription, the abstract motion-regularization term $\mathcal{J}_{\mathrm{m}}(\boldsymbol q)$ in P1 is instantiated by the squared first-order control-point differences and the squared second-order B\'ezier differences. Combining the LNF constraints, predicate-consistency constraints, B\'ezier motion constraints, serving-BS association, and handover indicators yields the following mixed-integer quadratically constrained reformulation.
\begin{equation*}
\begin{aligned}
	\mathrm{(P2)}\quad
	\min_{\mathcal{X}_{\mathrm{P2}}}\quad
	&\sum_{n=0}^{N-1}
	\Big(
	w_p\sum_{\ell=0}^{2}\|\boldsymbol r_{n,\ell}^{(1)}\|_2^2
	\\
	&\qquad
	+w_s\sum_{\ell=0}^{1}\|\boldsymbol d_{n,\ell}^{(2)}\|_2^2
	\Big)
	+w_{\mathrm{ho}}\sum_{n=0}^{N-2}\eta_n\\
	\mathrm{s.t.}\quad
	&\boldsymbol r_{0,0}=\boldsymbol q_{s,xy},\ 
	\eqref{eq:bezier_continuity}--\eqref{eq:soc_speed_bezier},\ 
	\eqref{eq:one_bs_segment},\ 
	\eqref{eq:coverage_big_m},\ 
	\eqref{eq:handover_binary},\\
	&\boldsymbol{\chi}^{\varphi}\in\mathcal{Z}_{\varphi}(\boldsymbol r),\ 
	\eqref{eq:lnf_edge_set}--\eqref{eq:lnf_flow},\ 
	\eqref{eq:bezier_second_difference}.
\end{aligned}
\end{equation*}
Here $\mathcal{X}_{\mathrm{P2}}=\{\boldsymbol r,\delta,\eta,\xi,\boldsymbol{\chi}^{\varphi},y,\boldsymbol\omega\}$ collects the decision variables. The handover weight $w_{\mathrm{ho}}$ follows P1, while $w_p$ and $w_s$ instantiate the abstract motion-regularization functional $\mathcal{J}_{\mathrm{m}}(\boldsymbol q)$ through first-order motion effort and second-order trajectory smoothness, respectively. The weights normalize the heterogeneous motion and handover terms and are selected empirically. The variable domains are $\delta_{n,i}\in\{0,1\}$, $\eta_n\in\{0,1\}$, $y_e\in\{0,1\}$, $\xi_{k,j,\ell,r}\in\{0,1\}$, $\boldsymbol{\chi}^{\varphi}\in\{0,1\}^{N_{\varphi}}$, and $\boldsymbol\omega_e\in[0,1]^{N_{\varphi}}$. In our implementation, P2 is solved in Gurobi as an MIQCP, where the norm constraints are represented as second-order cone constraints.

\begin{table}[!h]
	\centering
	\caption{Main Simulation Parameters}
	\begin{tabular}{lc}
		\hline
		\textbf{Parameter} & \textbf{Value} \\
		\hline
		Workspace & $5~\mathrm{km}\times5~\mathrm{km}$ \\
		BS map source & Baseline cellular layout \\
		Number of BS regions & $30$ \\
		Finite-blocklength $(B,\tau,n)$ & $(1~\mathrm{MHz},1~\mathrm{ms},1000)$ \\
		URLLC target $(\varepsilon_{\max},R_{\mathrm{req}})$ & $(10^{-5},0.25)$ \\
		UAV altitude & $H=300~\mathrm{m}$ \\
		Start position & $(250,2250,300)~\mathrm{m}$ \\
		B\'ezier segments & $N=24$ \\
		Nominal segment duration & $\Delta t=10~\mathrm{s}$ \\
		Fixed mission horizon & $T=240~\mathrm{s}$ \\
		Maximum speed & $v_{\max}=35~\mathrm{m/s}$ \\
		Weights $(w_p,w_s,w_{\mathrm{ho}})$ & $(0.5,0.005,10^4)$ \\
		\hline
	\end{tabular}
	\label{tab:sim_params_lnf_bezier}
\end{table}

\begin{table*}[t]
	\centering
	\caption{Mission Library Used in the Simulation}
	\small
	\setlength{\tabcolsep}{3pt}
	\begin{tabular}{lp{0.52\textwidth}p{0.33\textwidth}}
		\hline
		\textbf{Mission} & \textbf{STL specification} & \textbf{Mission meaning} \\
		\hline
		M1 &
		$\begin{gathered}
		\varphi_1=\Diamond_{[0,T]}\pi_A
		\wedge\Diamond_{[0,T]}\pi_B
		\wedge\Diamond_{[0,T]}\pi_R\\
		\wedge(\neg\pi_B\,\mathcal{U}_{[0,T]}\pi_A)
		\wedge(\neg\pi_R\,\mathcal{U}_{[0,T]}\pi_B)
		\end{gathered}$ &
		Visit $A$, then $B$, and finally $R$ within $[0,T]$. \\
		M2 &
		$\begin{gathered}
		\varphi_2=\Diamond_{[0,T]}\pi_A
		\wedge(\Diamond_{[0,T]}\pi_B\vee\Diamond_{[0,T]}\pi_C)
		\wedge\Diamond_{[0,T]}\pi_R\\
		\wedge(\neg\pi_B\,\mathcal{U}_{[0,T]}\pi_A)
		\wedge(\neg\pi_C\,\mathcal{U}_{[0,T]}\pi_A)\\
		\wedge\big((\neg\pi_R\,\mathcal{U}_{[0,T]}\pi_B)
		\vee(\neg\pi_R\,\mathcal{U}_{[0,T]}\pi_C)\big)
		\end{gathered}$ &
		Visit $A$ within $[0,T]$, then either $B$ or $C$ within $[0,T]$, and finally $R$ within $[0,T]$. \\
		M3 &
		$\begin{gathered}
		\varphi_3=\Diamond_{[0,40]}\pi_A
		\wedge\Diamond_{[90,150]}\pi_B
		\wedge\Diamond_{[210,T]}\pi_R\\
		\wedge(\neg\pi_B\,\mathcal{U}_{[0,T]}\pi_A)
		\wedge(\neg\pi_R\,\mathcal{U}_{[0,T]}\pi_B)
		\end{gathered}$ &
		Visit $A$ within $[0,40]$~s, inspect $B$ within $[90,150]$~s, and return to $R$ within $[210,T]$. \\
		M4 &
		$\begin{gathered}
		\varphi_4=\Diamond_{[0,60]}\pi_A
		\wedge\Diamond_{[90,150]}\pi_B
		\wedge\Diamond_{[180,220]}\pi_C
		\wedge\Diamond_{[220,T]}\pi_R\\
		\wedge(\neg\pi_B\,\mathcal{U}_{[0,T]}\pi_A)
		\wedge(\neg\pi_C\,\mathcal{U}_{[0,T]}\pi_B)
		\wedge(\neg\pi_R\,\mathcal{U}_{[0,T]}\pi_C)
		\end{gathered}$ &
		Visit $A$ within $[0,60]$~s, serve $B$ within $[90,150]$~s, upload data at $C$ within $[180,220]$~s, and return to $R$ within $[220,T]$. \\
		\hline
	\end{tabular}
	\label{tab:mission_library}
\end{table*}

\section{Numerical Results}
\subsection{Simulation Setup}

We evaluate the proposed planner in a cellular-connected UAV mission scenario over a $5~\mathrm{km}\times5~\mathrm{km}$ square area. The UAV flies at a fixed altitude $H=300~\mathrm{m}$ from the initial position $\boldsymbol{q}_s=[250,\,2250,\,300]^\top~\mathrm{m}$. The baseline cellular layout contains $M=30$ BS coverage disks generated from the finite-blocklength threshold. The main simulation parameters are summarized in Table~\ref{tab:sim_params_lnf_bezier}.

The serving margin reported in the simulations is computed from the selected disk and the B\'ezier control points. If $b_n$ is the BS selected for segment $n$, the segment margin is
\begin{equation}
\begin{aligned}
	m_n&=\rho_{b_n}-\max_{\ell=0,\ldots,3}
	\|\boldsymbol r_{n,\ell}-\boldsymbol c_{b_n}\|_2,\\
	m_{\min}&=\min_{n=0,\ldots,N-1}m_n.
\end{aligned}
	\label{eq:serving_margin}
\end{equation}
This definition is consistent with \eqref{eq:coverage_big_m}, since URLLC feasibility is imposed on all control points of the selected B\'ezier segment.

The numerical study is organized as follows. First, we evaluate mission generality over four STL specifications under the same network map and planner. Second, we study the handover-awareness tradeoff by sweeping the handover weight $w_{\mathrm{ho}}$ in M1. Third, we tighten the offline finite-blocklength coverage map by increasing $R_{\mathrm{req}}$ while keeping the calibrated BS powers fixed. Finally, we report the computational footprint of the LNF instances to show how predicate count and LNF size relate to solve time.

\subsection{Mission Generality Under STL Specifications}

Four semantic regions are used in the mission library, namely inspection region $A$, service region $B$, optional data-upload region $C$, and return region $R$. Their centers are $(1156.44,2515.16)$~m, $(3965.45,2363.05)$~m, $(3176.99,4635.40)$~m, and $(3963.76,4306.47)$~m, respectively. The axis-aligned square regions have half-side lengths $300$~m for $A$ and $B$, and $320$~m for $C$ and $R$. The four STL specifications are listed in Table~\ref{tab:mission_library}, where all temporal bounds are given in seconds and $T=240~\mathrm{s}$.

Table~\ref{tab:mission_generality_results} reports the results for M1--M4 under the same cellular map, B\'ezier motion model, objective weights, and computation budget. All four instances return feasible trajectories satisfying the encoded STL, motion, handover, and disk-based URLLC coverage constraints. Since each serving-BS region is enforced as a quadratic disk constraint rather than a polygonal inner approximation, the resulting mixed-integer quadratically constrained programs are demanding. Mission-specific warm starts are used to initialize the serving-cell sequence and improve incumbent discovery. The reported values compare the resulting mission timing, handover behavior, trajectory length, serving margin, and computational runtime across different STL specifications.

\begin{table}[t]
	\centering
	\caption{Mission Generality Results Under the Same Network Map and Planner}
	\footnotesize
	\setlength{\tabcolsep}{2.5pt}
	\begin{tabular}{lccccc}
		\hline
		\textbf{Mission} & $N_{\varphi}$ & \textbf{HO} & \textbf{Path (km)} & \textbf{Min. margin (m)} & \textbf{Time (s)} \\
		\hline
		M1 & $72$ & $7$ & $6.185$ & $1.933$ & $147.35$ \\
		M2 & $96$ & $8$ & $5.030$ & $0.046$ & $127.41$ \\
		M3 & $72$ & $7$ & $4.949$ & $0.006$ & $98.20$ \\
		M4 & $96$ & $6$ & $5.443$ & $0.045$ & $97.28$ \\
		\hline
		\multicolumn{6}{l}{\footnotesize $N_{\varphi}$ denotes the number of predicate-time variables.}
	\end{tabular}
	\label{tab:mission_generality_results}
\end{table}

The path lengths range from $4.949$~km to $6.185$~km across the mission library, and the handover counts range from six to eight. These aggregate results indicate that the proposed framework is sensitive not only to geometric path length but also to the interaction between STL mission phases and BS association. The trajectory of each mission is shown below with its corresponding analysis.

\begin{figure}[H]
	\centering
	\includegraphics[width=0.75\columnwidth]{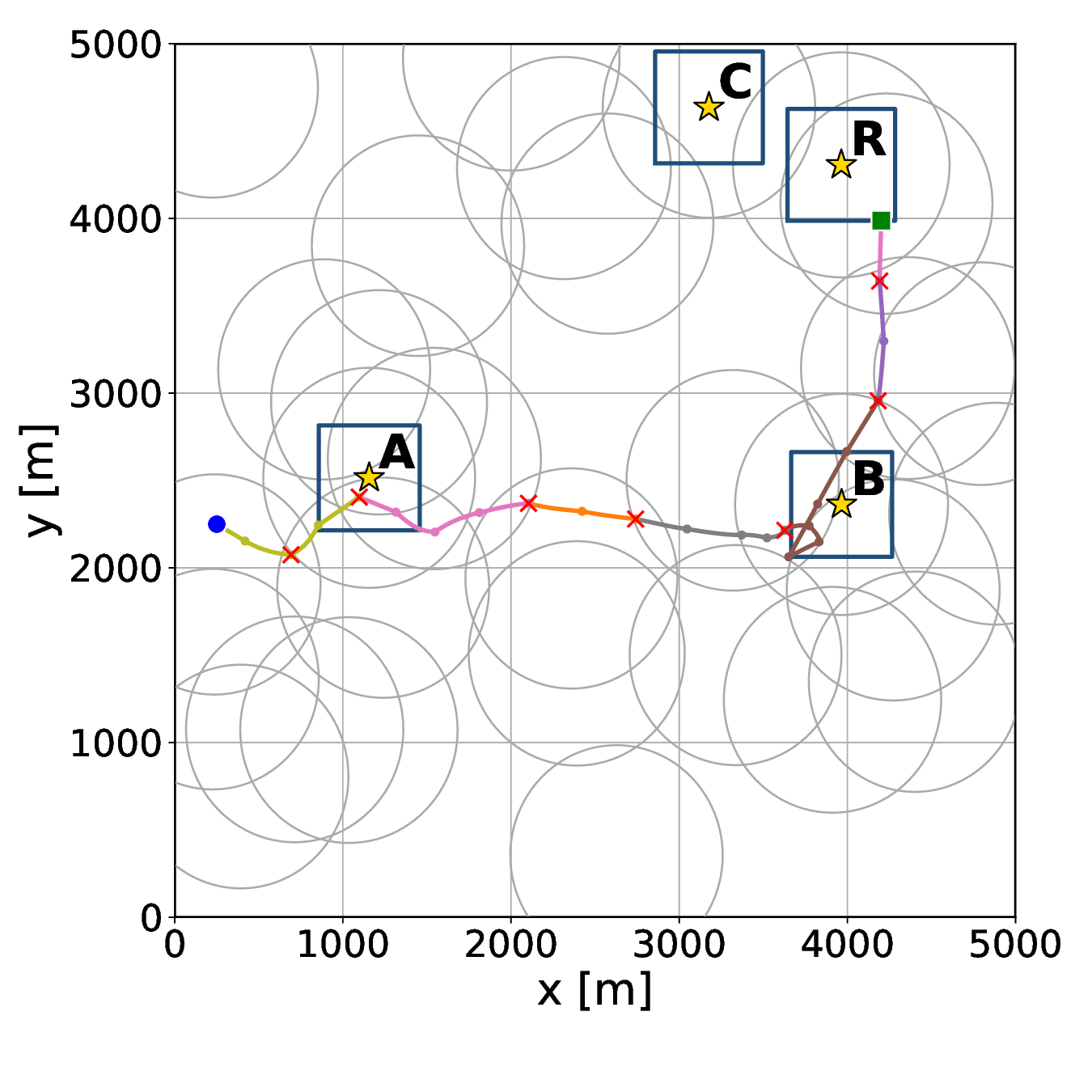}
	\vspace{-6pt}
	\caption{Optimized trajectory for M1 with ordered inspection and return.}
	\label{fig:mission_m1_trajectory}
\end{figure}

Fig.~\ref{fig:mission_m1_trajectory} shows the baseline ordered mission. The UAV visits $A$, then $B$, and finally returns to $R$ while maintaining URLLC feasibility over the selected BS sequence. This mission yields the longest path in Table~\ref{tab:mission_generality_results}, because the temporal ordering forces the route to traverse the inspection and service regions before returning.

\begin{figure}[H]
	\centering
	\includegraphics[width=0.75\columnwidth]{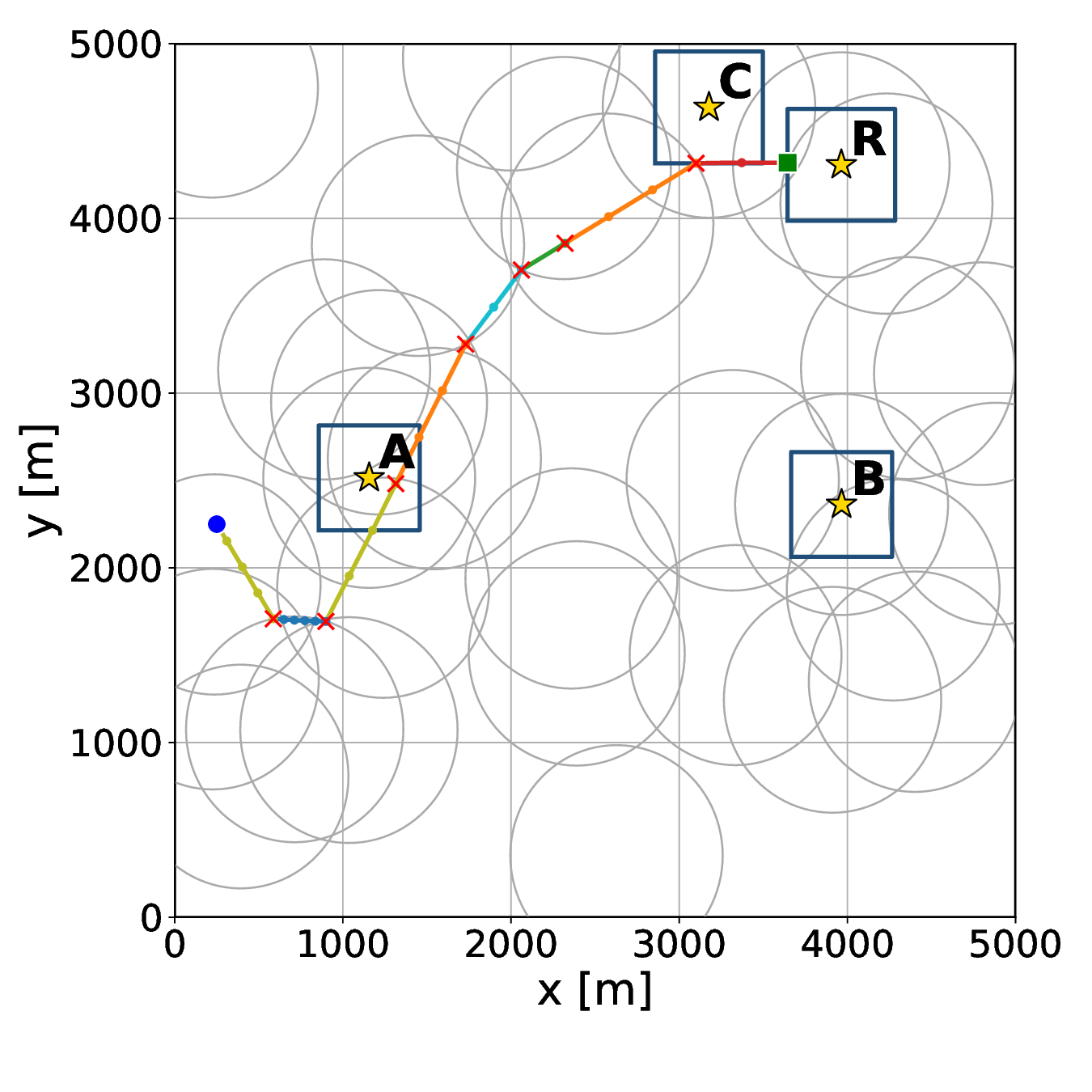}
	\vspace{-6pt}
	\caption{Optimized trajectory for M2 with disjunctive service selection.}
	\label{fig:mission_m2_trajectory}
\end{figure}

Fig.~\ref{fig:mission_m2_trajectory} corresponds to the disjunctive mission in which the UAV must visit $A$ and then choose either $B$ or $C$ before returning. Under the nominal handover-aware objective, the optimized trajectory satisfies the $C$ branch. The resulting serving margin is small, which indicates that the selected disjunctive branch is shaped by both mission timing and the boundary of the URLLC disk map.

\begin{figure}[H]
	\centering
	\includegraphics[width=0.75\columnwidth]{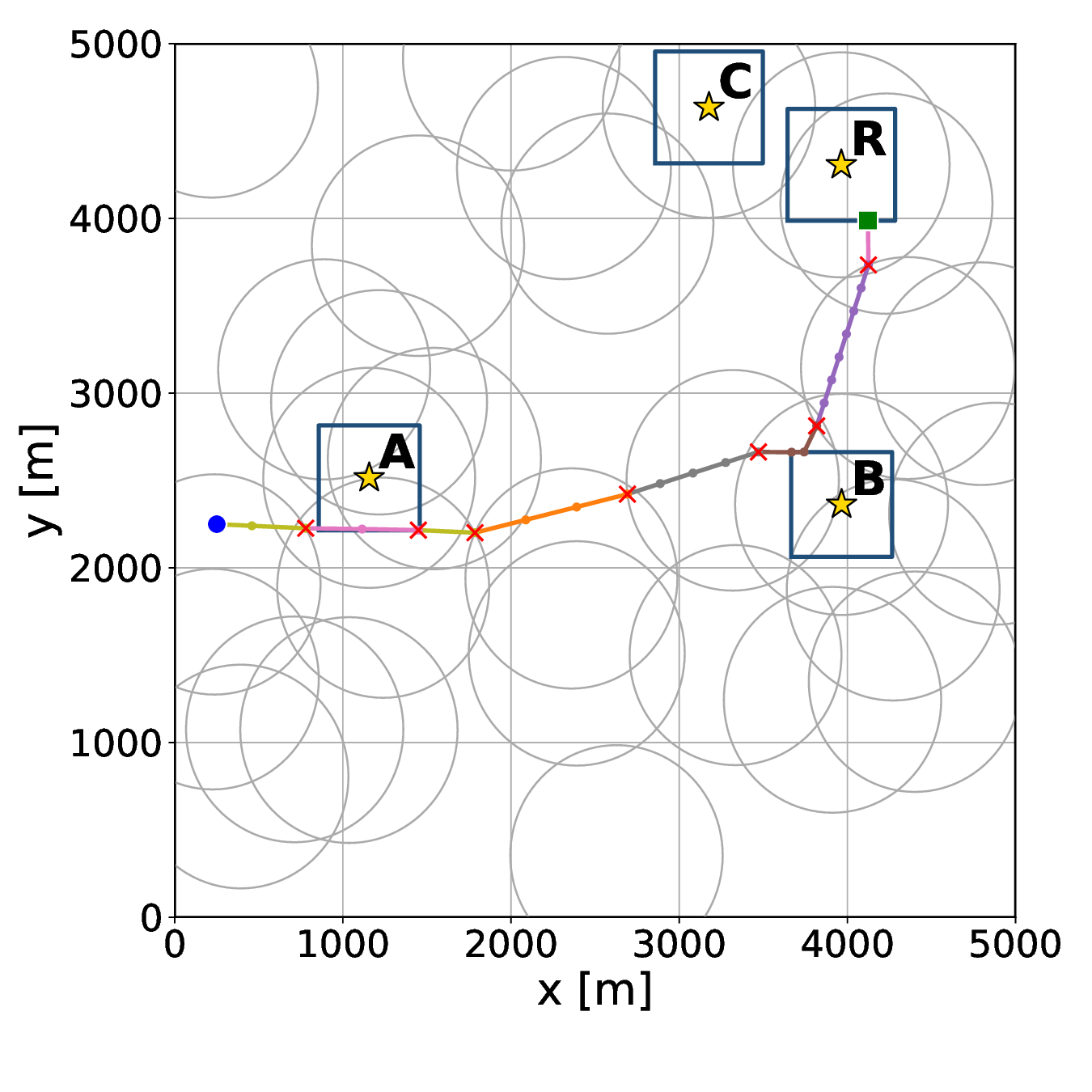}
	\vspace{-6pt}
	\caption{Optimized trajectory for M3 with deadline-constrained inspection.}
	\label{fig:mission_m3_trajectory}
\end{figure}

Fig.~\ref{fig:mission_m3_trajectory} illustrates the deadline-constrained inspection case. The early window for $A$, the scheduled window for $B$, and the late return window jointly restrict where the UAV can be at each segment-level time index. The path is shorter than M1 because the time windows guide the trajectory toward a more direct mission sequence, but the minimum serving margin is also the smallest among the four missions.

\begin{figure}[H]
	\centering
	\includegraphics[width=0.75\columnwidth]{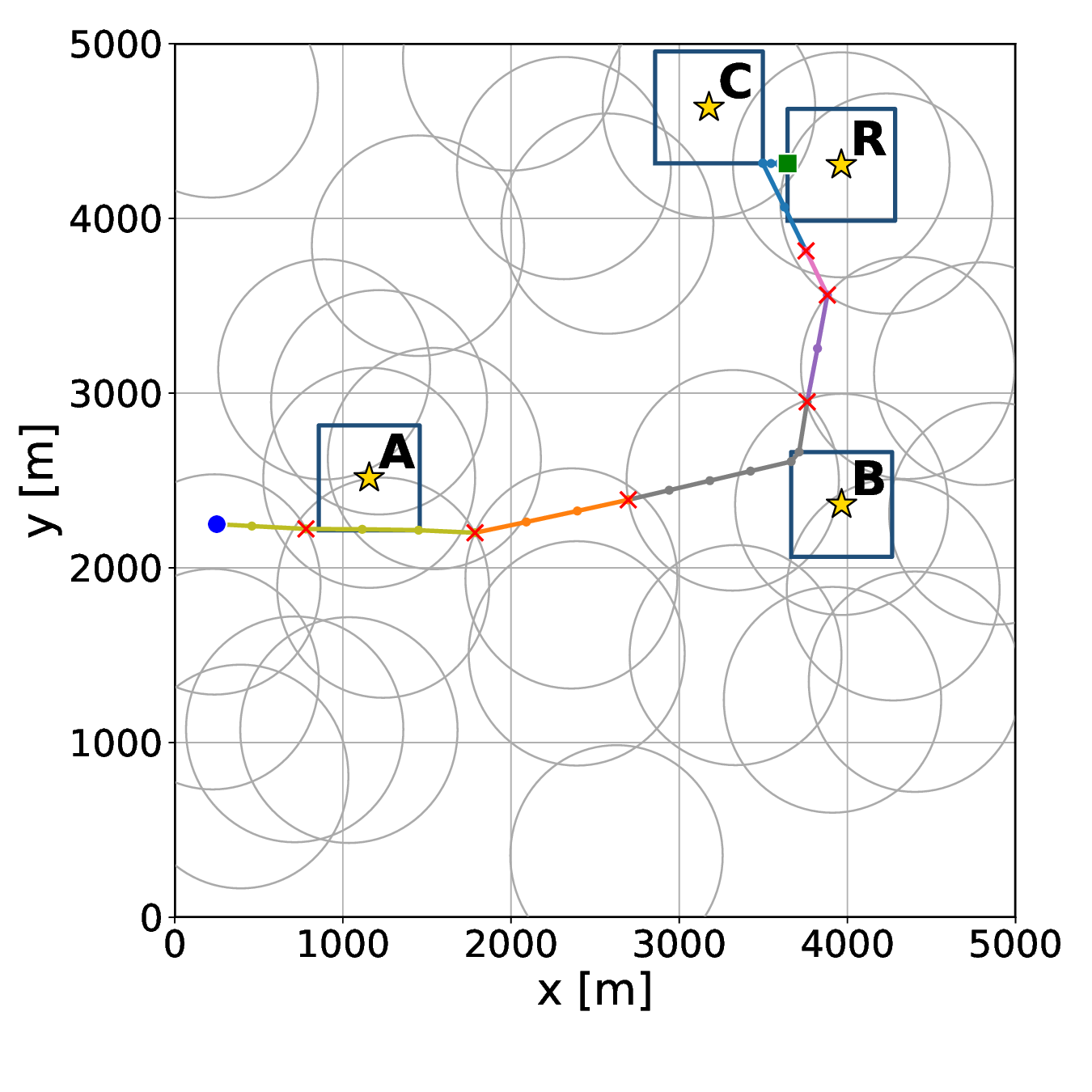}
	\vspace{-6pt}
	\caption{Optimized trajectory for M4 with data-upload service.}
	\label{fig:mission_m4_trajectory}
\end{figure}

Fig.~\ref{fig:mission_m4_trajectory} adds a data-upload visit to $C$ after the service phase at $B$. Compared with the ordered mission in M1, this extra mission phase changes the route and serving-cell sequence, but the handover count is reduced to six because the optimizer can maintain several consecutive segments under the same serving BS while moving toward $C$ and $R$.

\subsection{Handover Awareness and Finite-Blocklength Stringency}

The handover-weight sensitivity study is conducted based on M1. For $w_{\mathrm{ho}}=10^3$, $5\times10^3$, and $10^4$, the handover counts are seven, six, and seven, respectively, while the minimum serving margins are approximately $0$~m, $0$~m, and $1.933$~m. These results indicate that changing $w_{\mathrm{ho}}$ can alter the serving-BS sequence and serving margin, but the handover count is not monotone because the selected mission is also constrained by the overlap between the required mission regions and the URLLC disks. We therefore use this sweep as a sensitivity check rather than as the main evidence for a monotone handover-reduction trend.

\begin{figure}[h]
	\centering
	\includegraphics[width=0.7\columnwidth]{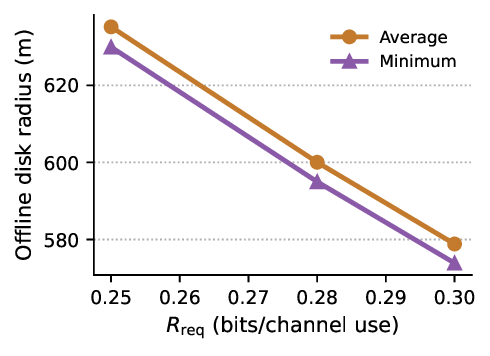}
	\vspace{-10pt}
	\caption{Average and minimum offline disk radii under different finite-blocklength rate requirements.}
	\label{fig:sensitivity_fbl_radius}
\end{figure}

\begin{figure}[h]
	\centering
	\includegraphics[width=0.7\columnwidth]{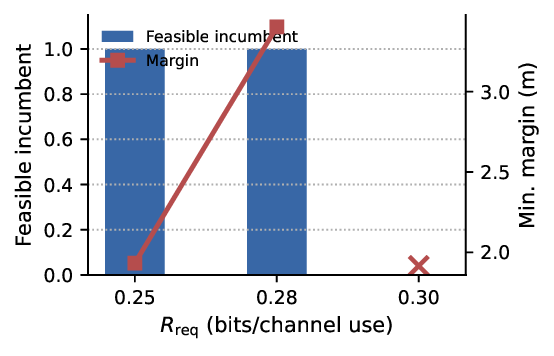}
	\vspace{-10pt}
	\caption{Feasible incumbent indicator and minimum serving margin under different finite-blocklength rate requirements.}
	\label{fig:sensitivity_fbl_feasibility}
\end{figure}

Figs.~\ref{fig:sensitivity_fbl_radius} and~\ref{fig:sensitivity_fbl_feasibility} evaluate finite-blocklength stringency. The offline disk map is first calibrated at $R_{\mathrm{req}}=0.25$ bits/channel use and then recomputed with the calibrated BS powers fixed. Increasing $R_{\mathrm{req}}$ from $0.25$ to $0.30$ raises the required SNR from $0.291$ to $0.342$ and reduces the average disk radius from $635.17$~m to $578.83$~m. The planner finds feasible incumbents for $R_{\mathrm{req}}=0.25$ and $0.28$, with minimum serving margins of $1.933$~m and $3.403$~m, respectively. Under the same $180$~s time limit, no feasible incumbent is found at $R_{\mathrm{req}}=0.30$. This does not prove mathematical infeasibility, but it shows that the current mission is close to the URLLC coverage boundary and that stricter finite-blocklength requirements can remove the effective planning margin.

\subsection{Computational Footprint}

\begin{figure}[h]
	\centering
	\includegraphics[width=0.8\columnwidth]{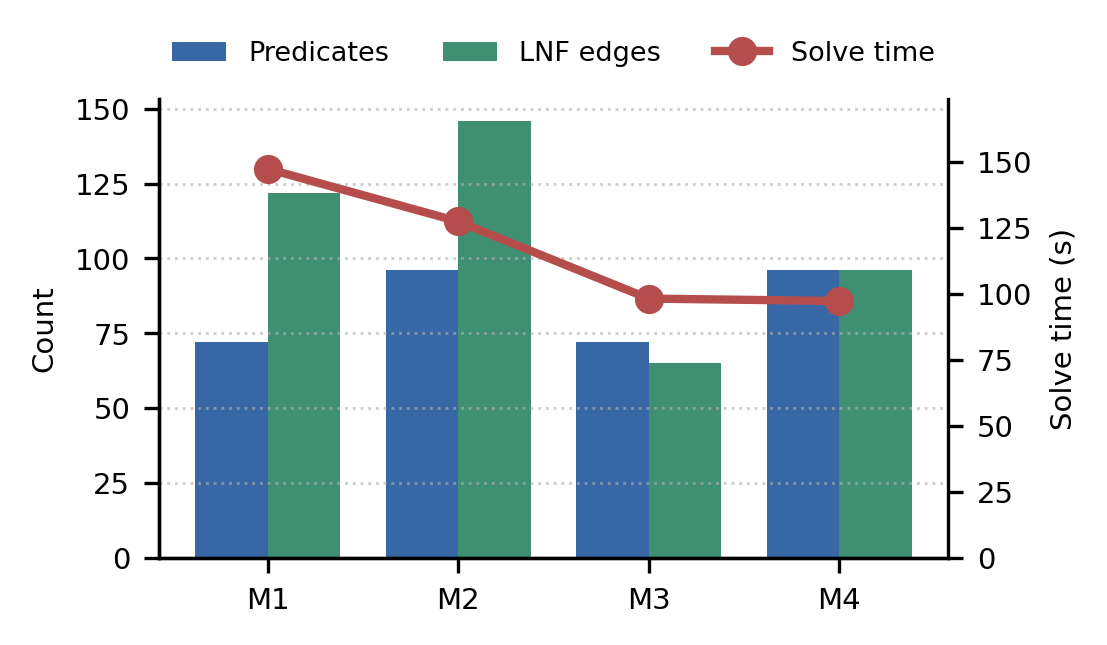}
	\vspace{-10pt}
	\caption{Computational footprint of the mission library. Predicate-time variable count, LNF edge count, and solve time capture complementary aspects of the coupled STL, handover, and B\'ezier planning problem.}
	\label{fig:complexity_runtime}
\end{figure}

Fig.~\ref{fig:complexity_runtime} compares the number of unique predicate-time variables, LNF edges, and solve times across the mission library. The LNF size reflects the temporal-logic structure of the mission. M2 grows due to the disjunctive branch, and M4 grows because it includes an additional data-upload region. Runtime, however, is not determined by the LNF size alone. The control-point-hull predicate encoding couples each logical predicate with multiple B\'ezier control points, so solve time also depends on the geometry of feasible mission regions and their overlap with URLLC disks. This confirms that the computational burden comes from the combined mission-network-motion structure rather than from the STL formula size alone.

\FloatBarrier
\section{Conclusion}
In this paper, we studied handover-aware trajectory planning for cellular-connected UAVs executing STL-specified missions under finite-blocklength URLLC constraints. We formulated a joint mission-network-motion optimization problem that couples STL satisfaction, B\'ezier-parameterized motion, BS association, and handover behavior, and transformed it into a mixed-integer quadratically constrained model through an LNF-based STL reformulation. Simulations on four STL missions show that the same planner can handle ordered inspection, disjunctive service selection, deadline-constrained inspection, and data-upload tasks under one network map. The results also show that mission timing and URLLC coverage jointly shape the feasible trajectory, serving margin, and handover sequence. Future work will extend the framework to time-varying wireless maps and scalable multi-UAV planning.

	\bibliographystyle{IEEEtran}
	\bibliography{reference}
	
\end{document}